\documentclass{emulateapj}

\newcommand{\Mwd}{\mbox{$M_{\mathrm{wd}}$}}
\newcommand{\Rwd}{\mbox{$R_{\mathrm{wd}}$}}
\newcommand{\Twd}{\mbox{$T_{\mathrm{wd}}$}}
\newcommand{\Tcent}{\mbox{$T_{\mathrm{cent}}$}}
\newcommand{\Oang}{\mbox{$\theta_{\rm spot}$}}
\newcommand{\Beta}{\mbox{$\beta_{\rm spot}$}}
\renewcommand{\Phi}{\mbox{$\phi_{\rm spot}$}}
\newcommand{\Teff}{\mbox{$T_{\mathrm{eff}}$}}
\newcommand{\La}{\mbox{${\mathrm{Ly\alpha}}$}}
\newcommand{\Lb}{\mbox{${\mathrm{Ly\beta}}$}}
\newcommand{\Lg}{\mbox{${\mathrm{Ly\gamma}}$}}

\newcommand{\Line}[3]{\Ion{#1}{#2}\,$\lambda$\,#3}
\newcommand{\Lines}[3]{\Ion{#1}{#2}\,$\lambda\lambda$\,#3}
\newcommand{\Ion}[2]{#1{\,\scriptsize #2}}
\newcommand{\pmag}{\mbox{$\phi_\mathrm{mag}$}}
\newcommand{\porb}{\mbox{$\phi_\mathrm{orb}$}}
\newcommand{\kms}{\mbox{$\mathrm{km\,s^{-1}}$}}
\newcommand{\Msun}{\mbox{$M_{\odot}$}}
\newcommand{\msy}{\mbox{$\mathrm{\Msun\,yr^{-1}}$}}
\newcommand{\es}{\mbox{$\mathrm{erg\;s^{-1}}$}}
\newcommand{\ecsa}{\mbox{$\mathrm{erg\;cm^{-2}s^{-1}\mbox{\AA}^{-1}}$}}

\begin{document}

\title{\textit{FUSE} and \textit{HST}/STIS far-ultraviolet
  observations of AM\,Herculis in an extended low
  state\altaffilmark{1}}

\author{Boris T. G\"ansicke}
\affil{Department of Physics, University of Warwick, Coventry CV4 9BU}
\email{Boris.Gaensicke@warwick.ac.uk}
\author{Knox S. Long}
\affil{Space Telescope Science Institute, 3700 San Martin Drive,
Baltimore, MD 21218, USA}
\email{long@stsci.edu}
\author{Martin A. Barstow}
\affil{Department of Physics and Astronomy, University of Leicester,
  University Road, Leicester LE1 7RH, UK}
\email{mab@star.le.ac.uk}

\author{Ivan Hubeny}
\affil{Department of Astronomy and Steward Observatory, University of
  Arizona, Tucson, AZ 85721, USA}
\email{hubeny@aegis.as.arizona.edu}


\altaffiltext{1}{Based on observations made with the NASA/ESA Hubble
Space Telescope, obtained at the Space Telescope Science Institute,
which is operated by the Association of Universities for Research in
Astronomy, Inc., under NASA contract NAS 5-26555, and on observations
made with the NASA-CNES-CSA Far Ultraviolet Spectroscopic
Explorer. FUSE is operated for NASA by the Johns Hopkins University
under NASA contract NAS5-32985. }

\begin{abstract}
We have obtained \textit{FUSE} and \textit{HST}/STIS time-resolved
spectroscopy of the polar AM\,Herculis during a deep low state. The
spectra are entirely dominated by the emission of the white dwarf.
Both the far-ultraviolet (FUV) flux as well as the spectral shape vary
substantially over the orbital period, with maximum flux occurring at
the same phase as during the high state. The variations are due to the
presence of a hot spot on the white dwarf, which we model
quantitatively. The white dwarf parameters can be determined from a
spectral fit to the faint phase data, when the hot spot is
self-eclipsed. Adopting the distance of $79\pm{8 \atop6}$\,pc
determined by Thorstensen, we find an effective temperature of
$19\,800\pm700$\,K and a mass of
$\Mwd=0.78\pm{0.12\atop0.17}$\,\Msun. The hot spot has a lower
temperature than during the high state, $\sim34\,000-40\,000$\,K, but
covers a similar area, $\sim10$\% of the white dwarf surface. Low
state \textit{FUSE} and STIS spectra taken during four different
epochs in 2002/3 show no variation of the FUV flux level or spectral
shape, implying that the white dwarf temperature and the hot spot
temperature, size, and location do not depend on the amount of time
the system has spent in the low state.  Possible explanations are
ongoing accretion at a low level, or deep heating~--~both alternatives
have some weaknesses that we discuss.  No photospheric metal
absorption lines are detected in the \textit{FUSE} and STIS spectra,
suggesting that the average metal abundances in the white dwarf
atmosphere are lower than $\sim10^{-3}$ times their solar values.
\end{abstract}

\section{Introduction}
Polars, also known as AM\,Herculis stars, are a class of
cataclysmic variables which contain a magnetic white dwarf with
$B\ga10$\,MG. The strong magnetic field suppresses the formation of
an accretion disk, and channels the accreting material to small
regions near one or both of the white dwarfs magnetic poles. The
accreting material reaches the white dwarf with supersonic velocities,
is decelerated and heated in a shock and subsequently cools through
the emission of thermal X-rays and/or cyclotron radiation. Early
models \citep{lamb+masters79-1, king+lasota79-1} suggested that the
shock would stand above the white dwarf surface, and that
roughly half of the post-shock emission would be
intercepted by the white dwarf surface, heating a more or less
extended region below and around the shock to a few $10^5$\,K.
The prediction of these models was that the reprocessed
radiation will be observed in the soft X-ray band, and that the
luminosity of this reprocessed component should be roughly equal to the
sum of the observed luminosities in thermal bremsstrahlung and
cyclotron radiation.
Observationally, polars show a wide range of hard to soft X-ray
luminosity ratios, and some show a substantial excess of soft X-ray
emission compared to the predictions of the simple reprocessing model
(see \citealt{beuermann+burwitz95-1, ramsayetal94-1, ramsayetal96-1,
ramsay+cropper04-1} for references and discussion).  Detailed
hydrodynamic models show that the exact energy balance of the
stand-off shock depends on the magnetic field strength and the mass
flow rate \citep{fischer+beuermann01-1}, largely explaining the
deficiencies of the early models.  In the case of very high mass flow
rates, the shock may also be submerged in the white dwarf photosphere,
where the primary thermal bremsstrahlung is directly reprocessed into
soft X-ray emission, and hence hard X-rays are not observed at all
\citep{kuijpers+pringle82-1}.

A flaw in the initial reprocessing model was that, depending on the
shock height, the post-shock radiation may reach a fairly large area
on the white dwarf~--~especially so the cyclotron radiation which is
significantly beamed perpendicular to the magnetic field
lines. Distributing the post-shock radiation over a larger area would
result in a lower temperature in the heated region, shifting the peak
of the reprocessed emission to longer
wavelengths. \textit{International Ultraviolet Explorer}
(\textit{IUE}) ultraviolet observations of the bright prototype
AM\,Herculis revealed the presence of a rather large moderately hot
region on the white dwarf, first noticed by
\citet{heise+verbunt88-1}. Phase-resolved \textit{IUE} observations of
AM\,Her showed a quasi-sinusoidal flux modulation which
\citet{gaensickeetal95-1} explained by the presence of a hot spot
covering $\sim10$\% of the white dwarf surface, located near the
accreting magnetic pole. Comparing the luminosity of this pole cap,
\citet{gaensickeetal95-1} suggested that the original reprocessing
model may be valid in AM\,Her, but that the reprocessed component is
emitted in the ultraviolet (UV) rather than in soft X-rays.

A \textit{Hubble Space Telescope} (\textit{HST})/GHRS study of AM\,Her
in the high state confirmed the earlier \textit{IUE} results and
established a much better constraint on the size, location and
temperature of the pole cap \citep{gaensickeetal98-2}.  In order to
carry out a comparable high-quality study of the low state, we
obtained Directors Discretionary \textit{Far Ultraviolet Spectroscopic
Explorer} (\textit{FUSE}) and \textit{HST} observations of AM\,Her in
2002.

\begin{deluxetable}{lcccc}
\tablecolumns{5}
\tablewidth{0pc}
\tablecaption{\label{t-obslog}Log of the \textit{HST}/STIS and
  \textit{FUSE} observations of AM\,Her. For  the STIS observations,
  the magnitude from the associated acquisition image taken with the
  F28$\times$50LP is also reported.}
\tablehead{
\colhead{Dataset} &
\colhead{UT start} &
\colhead{Exp. time} &
\colhead{$m_\mathrm{acq}$}
}
\startdata
\textit{HST} \\
o8h801010 & 2002-07-11 12:18:14 & 2450s & 14.7\\
o8h801020 & 2002-07-11 13:35:40 & 3050s & --\\
o8h802010 & 2002-07-12 13:57:09 & 2450s & 14.6\\
o8h802020 & 2002-07-12 15:13:52 & 3030s & --\\
o8qp02010 & 2003-11-06 19:29:30 & 2200s & 14.6\\
o8qp03010 & 2003-11-07 09:44:41 & 2200s & 14.7\\
o8qp04010 & 2003-11-07 17:52:58 & 2200s & 14.6\\
o8qp07010 & 2003-11-08 09:44:15 & 2200s & 14.1 \\
\textit{FUSE} \\
P1840601001 & 2000-06-12 23:48 &  5643s & -- \\
Z0060101000 & 2002-05-11 23:39 & 46115s & -- \\
C0530503000 & 2002-09-08 05:51 &  4456s & -- \\
C0530504000 & 2002-09-08 09:12 &  7029s & -- \\
C0530505000 & 2002-09-08 14:12 &  5847s & -- \\
\enddata
\end{deluxetable}

\begin{figure}
\includegraphics[width=\columnwidth]{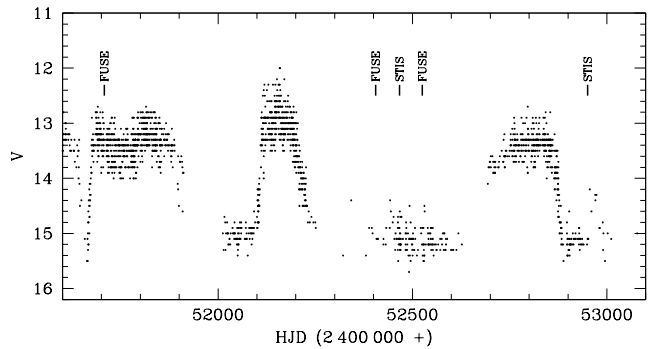}
\caption{\label{f-aavso} Long-term light curve of AM\,Her obtained by
  the AAVSO. The times of the three  FUSE and two \textit{HST}/STIS  observations
  discussed in this paper
 are indicated.}
\end{figure}

\section{Observations and Data Processing}

\subsection{HST/STIS}
\label{s-obshst} Time-resolved \textit{HST}/Space Telescope Imaging
Spectrograph (STIS) far-ultraviolet (FUV) spectroscopy of AM\,Her was
carried out during four spacecraft orbits on 11/12 July 2002
(Table\,\ref{t-obslog}).  At the time of the STIS observations AM\,Her
had been in the low state ($V\simeq15.2$) for $\sim200$\,d
(Fig.\,\ref{f-aavso}).  Because the orbital periods of AM\,Her and HST
are closely commensurate (185.7\,min and 96\,min, respectively) the
observations had to be scheduled in two blocks of two consecutive
orbits separated by a gap of 14 orbits to achieve full orbital phase
coverage. The data were taken using the G140L grating and the
$52\arcsec\times0.2\arcsec$ aperture, providing a spectral resolution
of $R\sim300$\,\kms\ over the wavelength range 1150$-$1715\,\AA, and
using the TIME-TAG acquisition mode which allows arbitrary temporal
binning of the data during the analysis.

AM\,Her was observed again with HST/STIS on 6/7/8 November 2003 over 4
\textit{HST} orbits as part of the program 9852 (PI Saar), using the
same setup as before except for the data acquisition mode
(Table\,\ref{t-obslog}).  The data were obtained in the ACCUM mode, in
which all registered photons are accumulated over a set exposure time
and then downlinked in form of a raw detector image. The observations
of AM\,Her were taken with an exposure time of 440s, resulting in a
total of 20 individual spectra. In November 2003, AM\,Her was observed when it
had been in a low state for $\sim60$\,d, following a $\sim180$\,d high
state, and just very shortly before it was rising to a short
($\sim30$\,d) intermediate state (Fig.\,\ref{f-aavso}).

\begin{figure*}
\includegraphics[angle=-90,width=\textwidth]{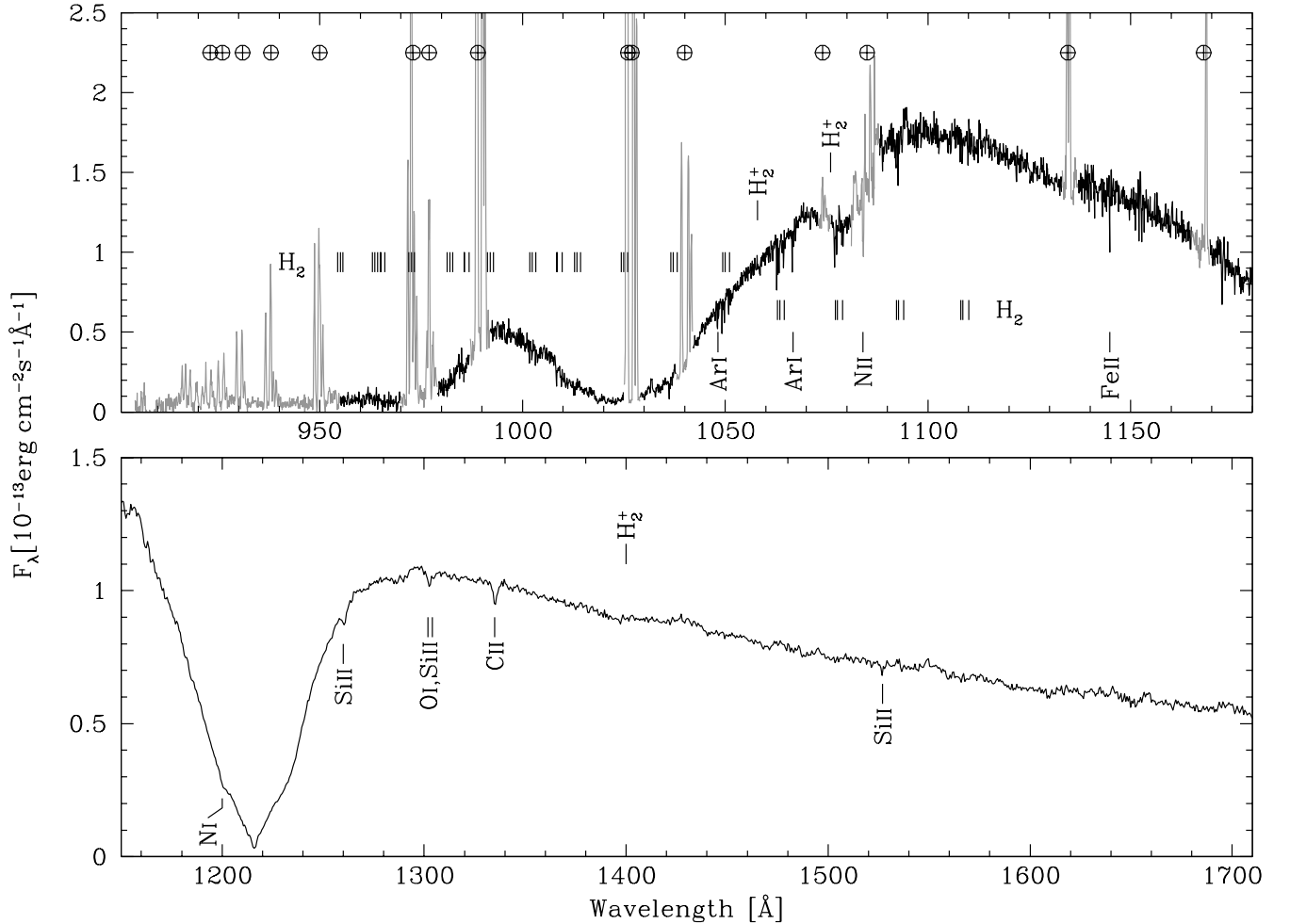}
\caption{\label{f-ave_spec} The grand average of the May 2002
  \textit{FUSE} spectra (top panel) and the July 2002
  \textit{HST}/STIS spectra (bottom panel) of
  AM\,Herculis. \textit{FUSE} regions contaminated by airglow are
  plotted in gray. Metallic interstellar absorption lines are
  indicated below the spectra, the positions of interstellar $H_2$
  molecular absorption bands are also given for the \textit{FUSE}
  range. Photospheric features from the white dwarf in AM\,Her are
   \La, \Lb, and \Lg, as well as weak absorption of $H_2^+$ at
  1400\,\AA, and more noticeable at 1076\,\AA. The 1058\,\AA\ $H_2^+$
  absorption line seen in single DA white dwarfs is not detected.}
\end{figure*}

We have used the STIS acquisition images to determine the brightness
of AM\,Her during the times of the FUV spectroscopy. The acquisition
images were  taken with the F28$\times$50LP filter,
which extends from 5400--10\,000\,\AA, with a pivot wavelength at
7229\,\AA, and compares in response roughly to an $R$-band filter.
The magnitudes reported in Table\,\ref{t-obslog} are entirely
consistent with AM\,Her being in a deep low state, except the last
data set, where $m_\mathrm{acq}\simeq14.1$. It may be that this
reflects the rising activity seen in the AAVSO light curve
(Fig.\,\ref{f-aavso}).

While inspecting the pipeline-calibrated STIS spectra and attempting a
first qualitative fit of the data using white dwarf model spectra (see
Sect.\,\ref{s-wdproperties} for details on the full model fits), we
noticed a flux deficit in the observed spectra at the blue end
($\lambda\la1170$\,\AA) of the G140L wavelength range. It is
well-known that the sensitivity of the STIS+G140L configuration is
time-variable, most noticeably a loss in sensitivity below
$\simeq1200$\,\AA, and is allegedly accounted for and corrected by the
CALSTIS pipeline. Intrigued by the apparent flux deficit, we retrieved
G140L spectra of the \textit{HST} flux standard Grw$+70^\circ$5824, a
$\Twd=20\,000$\,K DA white dwarf, taken on 12 August 1997 (shortly
after the commissioning of STIS) and on 3 August 2002, shortly after
our AM\,Her observations. Both spectra were reduced with CALSTIS V2.16
and the most recent reference files. Fitting the two STIS spectra of
Grw$+70^\circ$5824 it became clear that the flux calibration from the
pipeline does indeed underestimate the flux at the bluest wavelengths
in the more recent spectrum. We have therefore implemented the
following procedure to correct the time-dependent change in
sensitivity of the G140L grating that can be applied to observations
taken at any point throughout the life time of STIS. First, the
earliest spectrum of Grw$+70^\circ$5824 and a second spectrum taken
closest in time to the observation of the actual science target were
obtained from the \textit{HST} archive and pipeline-processed with the
time-dependent sensitivity correction switched off. Next, the two
spectra were binned in 3\,\AA\ steps, and the flux ratio
initial/recent is computed. This ratio is then smoothed with a 3-point
box car and used as a multiplicative correction for the target
spectrum.

\begin{figure}
\includegraphics[width=\columnwidth]{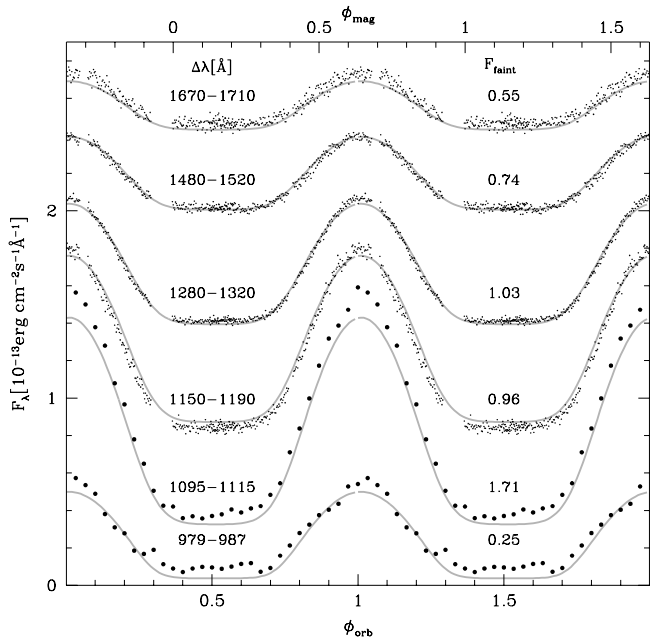}
\caption{\label{f-lightcurves} Small dots: phase-folded FUV light
curves created from the \textit{HST}/STIS time-tagged data at a time
resolution of 30\,s. The four curves were extracted over the
wavelength bands indicated on the left hand side, and offset in flux
by appropriate amounts. The faint-phase flux of each light curve is
given on the right hand side. Big dots: light curve created from
averaging the phase-resolved \textit{FUSE}  spectra
with a time resolution is 371\,s. Gray lines: best-fit white dwarf
plus hot spot model described in Sect.\,\ref{s-hotspot}.}
\end{figure}

\subsection{FUSE}
As shown in Figure \ref{f-aavso}, AM\,Her was observed with
\textit{FUSE} in May 2002 (our DDT program) and in September 2002 (our
additional GO program), almost exactly 60d before and after the July
2002 \textit{HST}/STIS observation. For completeness, we also include
the high-state observation carried out in June 2000
\citep{hutchingsetal02-1}. The \textit{FUSE} spectrograph consists
four independent optical channels that combined cover the 905-1187
\AA\ wavelength range \citep{moosetal00-1}. The optics of two of the
four channels are optimized for shorter wavelengths (905--1105\AA)
with a SiC coating. The optics of the other two channels are coated
with LiF and optimized for the longer wavelengths (1000--1187\AA). The
data are recorded in eight segments, A and B for each of 4 channels,
and the eight segments can be combined to cover the full
905--1187\,\AA\ range with some overlap. Both observations were taken
in the photon-counting time tag mode through the large
$30\arcsec\times30\arcsec$ (LWRS) aperture. This minimizes slit losses
that can occur due to misalignments of the four \textit{FUSE}
channels. \cite{sahnowetal00-1} describe the \textit{FUSE} observatory
and its in-flight performance in detail.

For the analysis here, the data were reprocessed using version 3.0 of
the CALFUSE software, using calibration files available in spring
2005.  Inspection of the data products indicate that both observations
were nominal and that slit losses were in fact quite small (less than
10\%).  As a result, it was not necessary to renormalize the spectra
from individual channels to create the combined spectra, either in the
time-averaged spectrum or in the phase resolved spectra discussed
below.  In generating the combined spectra, we rebinned the
data to 0.1\,\AA\ and weighted the various spectral channels
according to the effective area and exposure time for that particular
channel and wavelength.

\subsection{Average Spectra}\label{s-ave_spec}
As will be discussed in Sect.\,\ref{s-comparison}, the two sets of
\textit{FUSE} and \textit{HST} low state spectra (see
Fig.\,\ref{f-aavso}, Table\,\ref{t-obslog}) are practically identical,
and we therefore pursue the following analysis on the two longest
observations, the May 2002 \textit{FUSE} data, and the July 2002
\textit{HST} data. The average spectra calculated from these
observations are shown in Figure\,\ref{f-ave_spec}. The STIS spectrum
(bottom panel) is devoid of noticeable emission lines, confirming the
low accretion activity, and clearly reveals the broad photospheric
\La\ line of the white dwarf. The 1400\,\AA\ $H_2^+$ quasimolecular
absorption of \La\ is very weak, as expected for a temperature
$\ga20\,000$\,K. Weak absorption lines of CNO and Si are detected in
the spectrum, with equivalent widths of $\simeq100-200$\,m\AA. The
same transitions were detected at similar strengths in
\textit{HST}/GHRS high state data of AM\,Her
\citep{gaensickeetal98-2}. Based on the coincidence of the set of
detected transitions with the strongest interstellar absorption lines,
their low and apparently constant equivalent widths, and the fact that
only the resonance (ground-state) transitions (1260\,\AA, 1527\,\AA)
of the \Lines{Si}{II}{1260/65} and \Lines{Si}{II}{1527/33} doublets are
observed, we identify these lines as being due to interstellar
absorption. A limit on the metal abundances in the white dwarf
photosphere of AM\,Her will be derived in Sect.\,\ref{s-abundances}.

The \textit{FUSE} spectrum seamlessly connects to the STIS data, and
reveals the broad photospheric \Lb\ and \Lg\ lines.  Clearly present
is the 1076\,\AA\ component of the $H_2^+$ \Lb\ line, however, the
1058\,\AA\ component, which is of noticeable strength in
$\Twd\simeq20\,000$\,K DA white dwarfs \citep{koesteretal98-1,
hebrard+moos03-1}, is not detected. The absence of this feature is
intriguing, as \citet{dupuisetal03-1} also noticed a much weaker
1058\,\AA\ $H_2^+$ absorption in the magnetic ($B=2.3$\,MG) white
dwarf PG\,1658+441. They argued that the disagreement in the case of
the massive ($\log g=9.32$) white dwarf PG\,1658+441 might be due to
the lack of appropriate line profile data for high-density plasma. The
same argument would not work for AM\,Her, as it has a mass typical for
single white dwarfs. The disagreement between the predicted and
observed \Lb\ quasimolecular $H_2^+$ lines in AM\,Her may suggest that
the magnetic field plays a role in the formation of these lines.
The continuum flux of AM\,Her below $\simeq970$\,\AA, is very weak and
strongly contaminated by airglow emission (see
\citealt{feldmanetal01-1} for details on the identification of airglow
lines).  Similar to the STIS spectrum, a number of weak and sharp
absorption features are detected, which we identify as interstellar
metallic absorption lines \citep{sembach99-1} and $H_2$ molecular
absorption bands \citep{jenkins+peimbert97-1}.

\subsection{High Time Resolution Light Curves}
\label{s-lightcurves} We have used the TIME-TAG data obtained in
July 2002 data to produce FUV high time resolution light curves. The
general procedure of extracting background-subtracted light curves
from G140L TIME-TAG data has been described by
\citet{gaensickeetal01-1}. We decided to create light curves binned in
30\,s in the four bands 1150--1190\,\AA, 1280--1320\,\AA ,
1480--1520\,\AA, and 1670--1710\,\AA, which sample the blue and red
wing of the \La\ line as well as two continuum bands, respectively.
In addition, we have created lower time resolution (371\,s) light
curves from the phase-resolved \textit{FUSE} spectra in the wavelength
ranges 979--987\,\AA\ and 1095--1115\,\AA, which are free from airglow
emission. Orbital phases were computed using Tapia's linear
polarization ephemeris as quoted by \citet{heise+verbunt88-1}. The
accumulated error in the phase is $\la0.003$, which is negligibly
small for the purposes of our study.  The zero point of this phase
convention is defined as the phase of maximum linear polarization,
which occurs when the angle between the line of sight and the magnetic
axis reaches its maximum value, and we will refer to this phase
convention as magnetic phase, \pmag. The magnetic phase and the
orbital phase \porb, where the phase zero is defined as the inferior
conjunction of the secondary star is given by $\porb=\pmag+0.367$
\citep{gaensickeetal98-2}. The phase-folded light curves
(Fig.\,\ref{f-lightcurves}) display a strong orbital flux modulation,
with maximum flux near $\porb\simeq1.0$ and a flat minimum extending
over $\porb\simeq0.4-0.6$. The amplitude of the modulation peaks in
the 1095--1115\,\AA\ band, and decrease both to shorter and longer
wavelengths.  Following \citet{gaensickeetal95-1, gaensickeetal98-2},
we interpret the FUV flux variability as being due to the changing
aspect of a hot polar cap on the rotating white dwarf. The flat part
of the FUV light curve corresponds to the times when the heated pole
cap is self-eclipsed by the body of the white dwarf.

\begin{figure}
\includegraphics[width=\columnwidth]{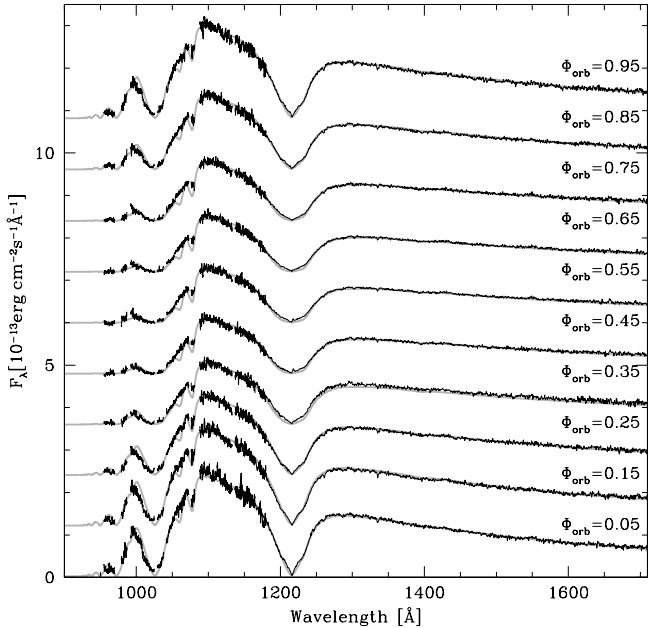}
\caption{\label{f-phase_spec} Phase-resolved \textit{FUSE} and STIS
  spectroscopy of AM\,Her (black), obtained in May 2002 and July 2002,
  respectively. The spectra are averages over 0.1 orbital phases and
  are centered on the orbital phases indicated in the plot. The
  bottom-most spectrum is plotted at its actual flux level, the other
  spectra are offset by 1.2 flux units each. Plotted in gray are the
  white dwarf plus hot spot spectra from the 3D model described in
  Sect.\,\ref{s-hotspot} for the corresponding phases.}
\end{figure}

\begin{figure}
\includegraphics[angle=-90,width=\columnwidth]{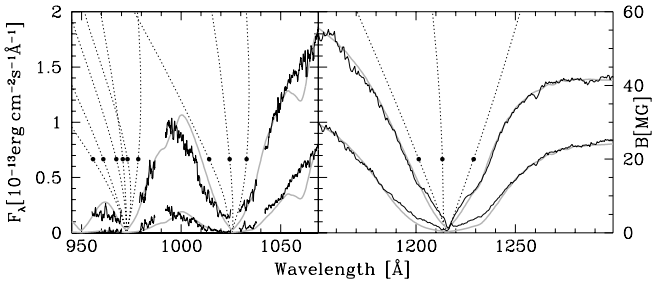}
\caption{\label{f-zeeman} \textit{FUSE} and \textit{HST}/STIS
spectroscopy of AM\,Her during the bright phase (\,=\,maximum
contribution of the hot spot, top curve) and faint phase
(\,=\,unheated white dwarf, bottom curve). The \textit{FUSE} spectra
have been binned to the same resolution as the STIS spectra. The
best-fit model spectra for the according phases (see
Sect.\,\ref{s-hotspot}) are plotted as thick gray lines.  Shown as
thin gray lines are the positions of the Zeeman components of \La,
\Lb, and \Lg\ as a function of magnetic field strength
\citep{friedrichetal96-3}. The black dots correspond to a field
strength of 20\,MG, as measured for the magnetic pole in AM\,Her
\citep{schmidtetal81-1}.}
\end{figure}

\subsection{Phase-resolved Spectra}
We have generated a total of 36 individual 300s exposures from the
July 2002 TIME-TAG data, corresponding to an orbital phase resolution
of $\simeq0.027$, which were then processed with the CALSTIS pipeline
within STSDAS. Orbital phases were computed as detailed in
Sect.\,\ref{s-lightcurves}. Finally, the 36 individual spectra were
averaged into 10 phase bins, resulting in effective exposure times of
600\,s to 1800\,s per phase bin.

The November 2003 STIS observations of AM\,Her were processed in an
analogous fashion  except that a
total of 20 spectra with exposures time of 440\,s each were used.

The process for generating phase resolved \textit{FUSE} spectra was
slightly different.  Basically the process was to concatenate all of
the data together into a individual raw TTAG data files for the
observation in May and September 2002. The data were then re-reduced
multiple times with CALFUSE, using "good time intervals" that
corresponded to each specific orbital phase, to produce fluxed
spectra in small (0.003 or 0.10) phase increments. As previously,
the individual channel spectra were co-added to produce spectra
covering the full spectral range.

Figure\,\ref{f-phase_spec} shows the May 2002 \textit{FUSE} and the
July 2002 \textit{HST}/STIS spectra averaged into 10 orbital phase
bins. Clearly noticeable is, in addition to the change in flux level,
the variation of the \La\ to \Lg\ line profiles which are narrowest
around $\porb\simeq1.0$ and broadest around $\porb\simeq0.5$.

\section{White dwarf properties}
\label{s-wdproperties}
We have created average STIS and FUSE spectra for the phase-range
where the hot polar cap is eclipsed, $0.40\la\porb\la0.60$
(Fig.\,\ref{f-lightcurves}), in order to investigate the properties of
the underlying white dwarf. As a preparation for the analysis carried
out in this and the following sections, we have generated a grid of
local thermal equilibrium (LTE) pure hydrogen (DA) non-magnetic white
dwarf models using the TLUSTY/SYNSPEC suite
\citep{hubeny88-1,hubeny+lanz95-1}. The white dwarf effective
temperatures and surface gravities covered were
$15\,000\,\mathrm{K}\le\Teff\le90\,000$\,K in appropriate steps,
$7.25\le\log g\le9.25$ in steps of 0.25, respectively. In the
temperature range considered here, especially for the unheated white
dwarf ($\sim20\,000$\,K, \citealt{heise+verbunt88-1,
gaensickeetal95-1}), NLTE primarily affects the line cores.  In the
case of AM\,Her, Zeeman splitting/broadening will be a more
substantial effect in the line cores. However, no self-consistent line
profile data for the case of combined Zeeman splitting and Stark
broadening is currently available, and hence no detailed model
spectra, are available. A quantitative modelling of the FUV data of
AM\,Her is therefore prone to some uncertainty, either by the use of
non-magnetic models as done in the present work, or by the use of
models with empirically weakened Stark broadening, as suggested by
\citet{jordan92-1}. The Zeeman splitting of the Lyman lines is
discussed below. The model spectra included the $H_2^+$ quasimolecular
lines of \La, \Lb, and \Lg, using the line profile data from
\citet{allardetal94-1, allardetal98-1, allardetal04-1}.

\subsection{Magnetic field}
\citet{schmidtetal81-1} detected several Zeeman components of the
white dwarf photospheric Balmer absorption lines in optical
spectropolarimetry obtained during a low state. From the positions of
these features, they deduced a magnetic field strength of
$B\sim10$\,MG at the magnetic equator and $B\sim20$\,MG at the
magnetic pole, assuming a simple dipole geometry of the field. An
independent measurement of the field strength was obtained from
near-infrared cyclotron emission, $B\simeq14.5$\,MG
\citep{baileyetal91-1}. While the Zeeman-splitting of the Lyman lines
is much weaker than that of the Balmer
lines \citep[e.g.][]{wunner87-1}, it may still have some noticeable
impact on the analysis of the FUV data of AM\,Her.

Figure\,\ref{f-zeeman} shows the orbital minimum and maximum
\textit{FUSE} and STIS spectra, corresponding to the undisturbed white
dwarf and the maximum flux contribution from the hot spot.  Two kinks
are apparent in the slope of the \La\ profile of the bright phase
spectrum, and their positions coincide with the $\sigma^{+/-}$ Zeeman
components of \La\ in a field strength of $\sim20$\,MG. This is
consistent with \citeauthor{schmidtetal81-1}'s results, as the
bright-phase spectrum is dominated by flux originating in the heated
pole cap near the magnetic pole. The observed splitting of
$\simeq15-20$\,\AA\ illustrates the effect of the magnetic field on
the \La\ line profile, and justifies the use of non-magnetic model
spectra for the analysis, as long as the core of \La\ is excluded from
the fits. No convincing features that could be associated with the
\La\ Zeeman components are seen in the faint phase spectrum, as the
\La\ profile is intrinsically too broad.

In the higher Lyman lines, the effect of Zeeman is more noticeable.
The \Lb\ $\sigma^{-}$ component might be responsible for the abrupt
change in slope seen in the blue wing of the bright phase \Lb\
profile. \Lg\ consists of 6 Zeeman components, and it is clear that
the use of non-magnetic models runs into more severe limitations for
the that line.

\begin{figure}
\includegraphics[angle=-90,width=\columnwidth]{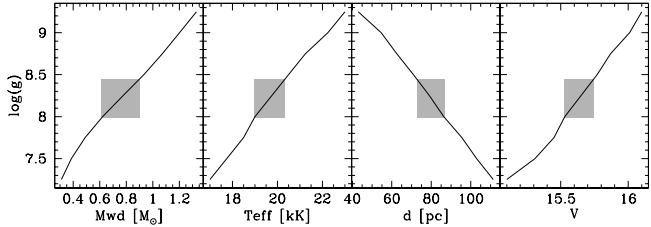}
\caption{\label{f-wdparameter} The white dwarf mass and temperature,
  distance to AM\,Her and the $V$-band magnitude of the white dwarf
  determined from fitting the STIS faint phase ($\porb=0.4-0.6$)
  spectrum. The gray shaded boxes correspond to the constraints
  imposed by \citeauthor{thorstensen03-1}'s (2003) astrometric
  distance measurement of $d=79\pm{8\atop6}$\,pc.}
\end{figure}

\subsection{Mass and temperature}
\label{s-wdmass}
We fitted the observed \textit{FUSE} and STIS faint phase spectra by stepping
through the model grid in $\log g$, leaving the effective temperature
and the flux scaling factor as free parameters. The flux scaling
factor is defined as $f/H=4\pi \Rwd^2/d^2$, where $f$, $H$, \Rwd, and
$d$ are the observed flux, the model flux, the white dwarf radius and
the distance to the system, respectively.  Assuming a \citet{wood95-1}
mass-radius relation for carbon-oxygen-core white dwarfs, the value of
$\log g$ defines both the white dwarf mass and the white dwarf radius.
Knowing $f/H$ and \Rwd, we then calculated the distance $d$. Hence,
our procedure results in best-fit values for \Twd, \Rwd, \Mwd, and $d$
as a function of $\log g$, as shown in Fig.\,\ref{f-wdparameter}. As
an additional control on the fit, we compute the $V$ magnitude of the
best-fit white dwarf models.  Increasing $\log g$ results in a higher
pressure in the white dwarf atmosphere, and therefore causes stronger
Stark broadening of the hydrogen lines. This effect is compensated by
an increased value of \Twd, which raises the degree of ionization of
hydrogen and narrows the hydrogen lines. The distance implied by the
fit decreases with increasing $\log g$ as a combined effect of \Rwd\
decreasing and \Twd\ increasing.

\begin{figure}
\includegraphics[angle=-90,width=\columnwidth]{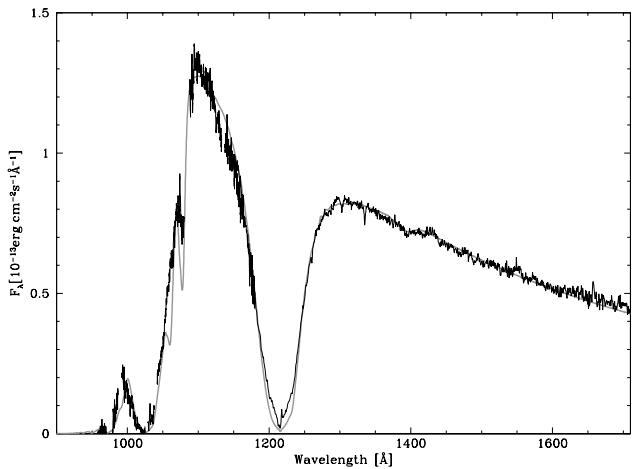}
\caption{\label{f-wdfaint}{Average \textit{FUSE} and STIS spectra
    (black) during the faint phase ($\porb=0.4-0.6$) when the hot spot
    is eclipsed by the body of the white dwarf. Shown in gray is the
    best-fit white dwarf model spectrum which is consistent with
    \citeauthor{thorstensen03-1}'s (2003) astrometric distance
    measurement of $d=79$\,pc. The most noticeable discrepancy between
    the model and the observed spectrum are found in the core of \La,
    in the quasi-molecular \Lb\ components, and in the blue wing of
    \Lg.}}
\end{figure}

As the distance to AM\,Her is fairly well established
\citep{thorstensen03-1}, $d=79\pm{8\atop6}$\,pc, the white dwarf mass
and temperature are constrained to the range where the distance
implied by the fit falls within the range of the astrometric parallax
measurement. The resulting ranges are $\Mwd=0.78\pm{0.12\atop0.17}$,
$\Rwd=7.61\pm{1.6\atop1.0}\times10^8$\,cm and $\Twd=19800\pm700$\,K,
where the errors are dominated by the remaining uncertainty in the
distance, and the best-fit model is shown in Fig\,\ref{f-wdfaint}. The
$V$ magnitude of the model is $\sim15.5-15.7$, consistent with
the observed low state magnitude of $V\simeq15.2$ which includes some
contribution from the secondary star. The
mass of the white dwarf in AM\,Her has been subject to substantial
debate, ranging from 0.39\,\Msun\ \citep{youngetal81-1} to
1.06\,\Msun\ \citep{cropperetal99-1}. The model-dependent
uncertainties of the method employed here (and previously in the case
of WZ\,Sge, \citealt{longetal04-1}) are small compared to the previous
estimates of the white dwarf mass in AM Her, and therefore we believe
our estimate of the mass is more accurate.  Indeed, the largest
uncertainty in the current estimate is due to the uncertainty in the
distance.

We note in passing that for single white dwarfs there is evidence that
their mean mass is higher than that of non-magnetic white dwarfs,
e.g. \citet{liebertetal03-1} quote $0.93$\,\Msun\ for magnetic white
dwarfs versus $0.6$\,\Msun\ for non-magnetic white dwarfs
\citep{bergeronetal92-1,finleyetal97-1,liebertetal05-1}. In CVs
white dwarf mass determinations are notoriously uncertain, and the
number of well-determined masses is yet too small to assess the
possibility of different white dwarf masses in magnetic and
non-magnetic CVs. In addition, white dwarf masses in CVs will be
subject to the details of the evolution of the system, i.e. depend
on the balance between mass accretion on one side, and mass ejection
during nova eruptions on the other side.

\subsection{Photospheric abundances}
\label{s-abundances}
Whereas in single white dwarfs gravitation separates the elements in
the envelope and results in most cases in either pure hydrogen (DA) or
pure helium (DB) atmospheres, it can be expected that in CVs accretion
of metal-rich material from the secondary star alters the chemical
composition of the white dwarf photosphere. In fact, FUV spectroscopy
of the white dwarfs in non-magnetic CVs has revealed substantial metal
abundances in all observed systems \citep[e.g.][]{sionetal90-1,
longetal93-1, gaensickeetal05-2}.

\begin{figure}
\includegraphics[width=\columnwidth]{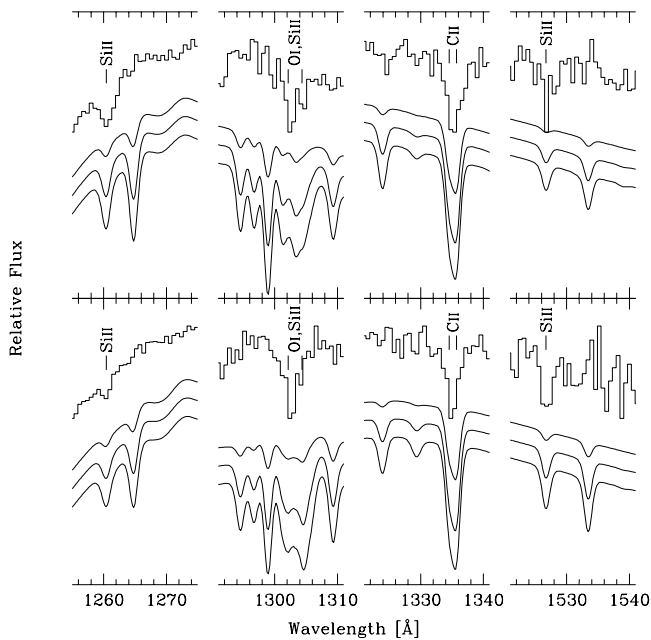}
\caption{\label{f-stis_abundances} \textit{HST}/STIS bright phase
  (top) and faint phase (bottom) spectra along with spectral models
  for the corresponding phases calculated for abundances at 0.001,
  0.005 and 0.01 times their solar values. The model
  spectra have been folded with the spectral response of the G140L
  grating, and are offset downwards by appropriate amounts. Strong
  interstellar absorption lines are indicated.}
\end{figure}

In polars, much less quantitative work on the photospheric abundances
of the accreting white dwarfs has been carried out.
\citet{gaensickeetal95-1} found two absorption lines near the
\Lines{Si}{II}{1260,65} doublet in nine out of eleven \textit{IUE} low
state spectra of AM\,Her and suggested that these structures may be
due to metals in the photosphere of the white dwarf.  However, they
also noticed that a broad absorption trough around 1300\,\AA\ (an
conglomerate \Ion{Si}{II}, \Ion{Si}{III}, and \Ion{O}{I} lines,
unresolved at \textit{IUE}'s resolution) was absent; this feature is
prominent in the \textit{IUE} spectra of e.g. VW\,Hyi, a dwarf nova
with a white dwarf of similar temperature to that in AM\,Her, and
therefore it was not entirely clear how to interpret the features that
were seen. \citet{depasquale+sion01-1} carried out spectral fits to
the \textit{IUE} data of AM\,Her that had been described  by
\citet{gaensickeetal95-1}, and derived metal abundances in the range
0.05 to 0.001, with no apparent correlation to either orbital phase
or the time spent in a low state.

\begin{figure}
\includegraphics[width=\columnwidth]{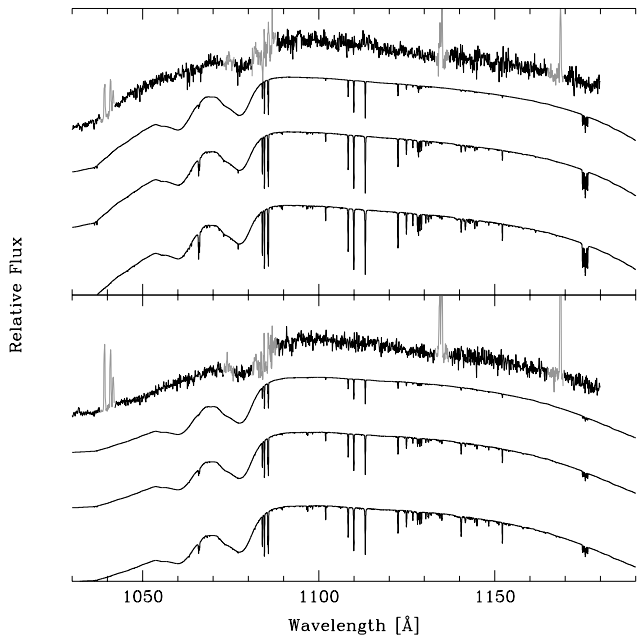}
\caption{\label{f-fuse_abundances} \textit{FUSE} bright phase (top)
  and faint phase (bottom) spectra along with spectral models for the
  corresponding phases calculated for abundances at 0.001, 0.005, and
  0.01 times their solar values. The model spectra have been offset
  downwards by appropriate amounts. Regions of the observed spectra
  affected by airglow are plotted in gray.}
\end{figure}

Figures\,\ref{f-stis_abundances} and \ref{f-fuse_abundances} show our
\textit{HST}/STIS and \textit{FUSE} spectra of AM\,Her, respectively,
at orbital maximum and minimum along with white dwarf model spectra
computed for the corresponding orbital phases assuming abundances of
0.001, 0.005, and 0.01 their solar values (and binned to the
corresponding spectral resolutions). As discussed in
Sect.\,\ref{s-ave_spec}, the noticeable absorption features in the
STIS spectra coincide with the strongest interstellar lines. 

One of the goals of our \textit{HST} and \textit{FUSE} low state
observations of AM\,Her was to detect photospheric metal lines and
probe into their Zeeman splitting at high field strengths.  To our
knowledge, no predictions for the Zeeman splitting of FUV metal
transitions at field strengths of tens of MG are available. Under the
linear Zeeman effect the surface field strengths in the range
$10-20$\,MG would result in Zeeman splitting of the order a few ten
\AA, depending on the Land\'e factor of the transition. However,
practically independent of the actual splitting, one would expect to
detect some noticeable absorption lines if substantial amounts of any
metal were present in the photosphere of the white dwarf. The absence
of any absorption features (except the Lyman lines) that could be
ascribed to photospheric absorption lines from the white dwarf
atmosphere strongly suggests that the average metal abundances over
the visibile hemisphere are lower than 0.001 times their solar values
at any given orbital phase.\footnote{A hypothetical explanation for
the absence of metals would be that mass transfer decreased to
extremely low levels, and that the metals diffused below the
photospheric level. However, this seems not likely, as only a minute
amount of accretion, $\sim10^{-15}$\,\msy, is necessary to enrich the
photosphere to a noticeable level, and observations both at X-ray,
optical, and IR wavelengths shows low-level accretion activity during the
low state \citep[e.g.][]{demartinoetal98-3, kafkaetal05-1,
baileyetal91-1}.}.

The finding of a nearly pure hydrogen composition of the atmosphere in
AM\,Her underlines fundamental differences in the accretion processes
in non-magnetic versus magnetic CVs.  \citet{beuermann+gaensicke03-1}
discussed that the high magnetic field strength in polars prohibits a
spreading of the accreting material lateral to the field lines down to
a depth in the atmosphere of the white dwarf where the gas pressure
equals the magnetic pressure. For typical white dwarf temperatures and
fields in polars, the decoupling occurs at several ten kilometer
depth, much below the observable photosphere. Enhanced metal
abundances could exist in the foot points of the accretion
column, but because of the small extent of these regions their
contribution to the average spectrum over the visible hemisphere is
probably negligible.

\begin{figure}
\includegraphics[width=\columnwidth]{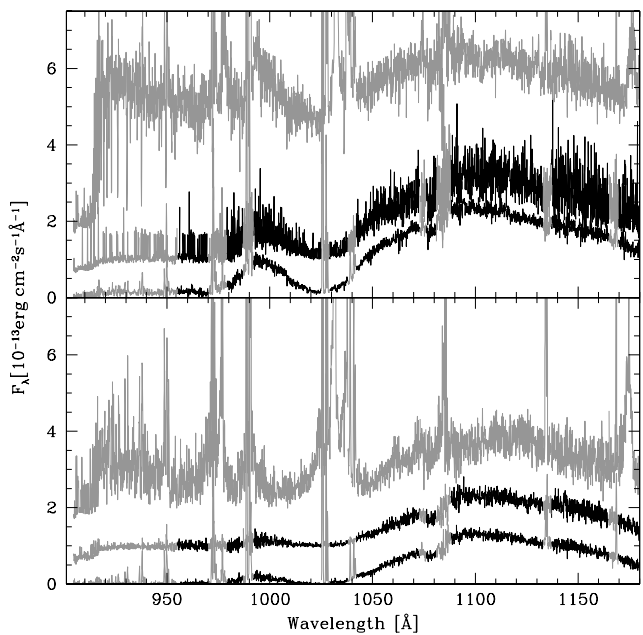}
\caption{\label{f-epoch_comp} Comparison between the \textit{FUSE}
observations of AM\,Her obtained at various epochs. The spectra
obtained at orbital maximum and minimum are plotted in the top and
bottom panels, respectively. In each of the panels, the bottom and
middle spectra are the the May 2002 (no flux offset) and the September
2002 (offset by one flux unit) low state observations. The top
spectrum, plotted in gray, is the June 2000 high state observation
(offset by two flux units, \citealt{hutchingsetal02-1}).}
\end{figure}

\section{The hot spot}
\label{s-hotspot} \citet{heise+verbunt88-1}  analyzed
\textit{IUE} spectroscopy of AM\,Her obtained during both high state
and low state. The authors concluded that the low state FUV emission
of AM\,Her is dominated by a $\simeq20\,000$\,K white dwarf, and that
during the high state a large area of the white dwarf is heated to
$\simeq30\,000$\,K. Using a substantially larger set of \textit{IUE}
spectra, \citet{gaensickeetal95-1} showed that during the low state
both the FUV flux and the shape of the \La\ absorption profile change
as a function of orbital phase, which they modeled in terms of a hot
polar cap with a temperature of $\simeq24\,000$\,K and a fractional
area of $\simeq0.08$ of the white dwarf surface, located near the
magnetic pole of the $\Twd\simeq20\,000$\,K white dwarf. The authors
showed that an analogous orbital modulation of the FUV flux, with the
same phases of minimum and maximum flux, is detected also during the
high state, and derived a pole cap temperature of $\simeq37\,000$\,K
and a similar fractional area as during the low state.

The analysis of the \textit{IUE} data was severely limited by the
low time resolution, implying substantial orbital phase smearing
over the course of the integration, the low
spectral resolution, and the large width of the geocoronal emission
\La\ line, effectively contaminating a substantial faction of the
white dwarf photospheric \La\ absorption.

\citet{gaensickeetal98-2} obtained high-time resolution (27s)
\textit{HST}/GHRS spectroscopy of AM\,Her in a high state covering the
range 1150--1435\,\AA\ at a spectral resolution of
$\simeq150$\,\kms. Continuum light curves showed a quasi-sinusoidal
modulation with a maximum near $\porb\simeq1.0$, and an amplitude
increasing towards shorter wavelengths. The authors fitted the light
curves with a three-dimensional white dwarf plus hot pole cap model,
where the temperature in the pole cap decreases from a central value
\Tcent\ to that of the unheated white dwarf, \Twd, as a linear
function of the opening angle \Oang. Fixing the distance to AM\,Her to
$d=90$\,pc and the binary inclination to $i=50^{\circ}$
\citet{gaensickeetal98-2} found $\Twd=21\,000$\,K,
$\Tcent=47\,000$\,K, $\Oang=69.2^{\circ}$ (corresponding to a
fractional white dwarf area of 9\%), $\Beta=54.4^{\circ}$ and
$\Psi=0.0^{\circ}$, where $\Beta$ is the colatitude and $\Psi$ the
azimuth of the spot.

We have analysed the \textit{FUSE} and STIS light curves described in
Sect.\,\ref{s-lightcurves} using the same three-dimensional model
described by \citet{gaensickeetal98-2}. However, in contrast to the
GHRS high state study, we fixed the white dwarf temperature and radius
to the values determined in Sect.\,\ref{s-wdmass} from the faint-phase
spectrum, ($\Twd=19\,800$, $\Rwd=7.61\times10^8$\,cm), as well as the
distance ($d=79$\,pc) and the binary inclination
($i=50^{\circ}$). Thus, we fit only \Tcent, \Oang, \Beta, and
$\Psi$. The best-fit for the combined set of \textit{FUSE} and STIS
light curves is obtained for $\Tcent=34\,000$\,K,
$\theta_\mathrm{spot}=82^{\circ}$ (corresponding to a fractional white
dwarf area of 12\%), $\psi_\mathrm{spot}=-4^{\circ}$,
$\beta_\mathrm{spot}=71^{\circ}$. There is some degeneracy in the spot
size and its temperature, allowing spot temperatures of up to
$\simeq40\,000$\,K (with $\Oang\simeq64^{\circ}$, corresponding to a
fractional white dwarf area of 8\%).

The best-fit model underestimates somewhat the 1095--1115\,\AA\
\textit{FUSE} continuum flux (Fig.\,\ref{f-lightcurves}). Given that
the absolute flux calibration of \textit{FUSE} is less well
established than that of STIS, we do not believe that adopting a more
complex model (meaning more free parameter) for the hot spot is
warranted by this small disagreement. However, despite the systematic
uncertainty between the instrumental calibrations, including the
\textit{FUSE} light curves into the fit does allow a tighter
constraint of \Tcent\ than fitting the STIS data alone. Interesting to
note is the slightly different shape of the \textit{FUSE} light curves
compared to the STIS data.  It appears that the flux maximum is
preceded by a small depression around $\porb\simeq0.9$, and that the
faint phase shows some rising trend between $0.4<\porb<0.6$. Again,
while this may point towards a somewhat more complex structure of the
hot spot compared with our simple circular model, our experience
suggests that \textit{FUSE} guiding is not sufficiently stable to
exclude an instrumental origin for these features. A simple constraint
on the geometry of the spot can be obtained by mirroring the STIS
light curves, and superimposing them to the original. The result is
that the original and the mirrored light curve are indistinguishable
over the entire orbital cycle, which implies a highly symmetric spot
with respect to the rotation axis.

Comparing the results for the low state to the numbers that
\citet{gaensickeetal98-2} found for the high state, the temperature of
the hot spot is obviously lower during the low state. On the other
hand, its size does not differ much, covering $\sim10$\% during the
low state and the high state. Similarly, the azimuth of the spot
during the low state does not change significantly between the low
state and the high state, although the low state co-latitude is higher
by $\sim15^{\circ}$ than for the high state. If the hot spot is due to
ongoing accretion (see Sect.\,\ref{s-discussion}), such a change in
\Beta\ could be related to a change in the location of the coupling
region (where the ballistic mass stream from the secondary couples to
the white dwarf magnetic field lines) as a result of the change in
mass transfer rate.

\begin{figure}
\includegraphics[width=\columnwidth]{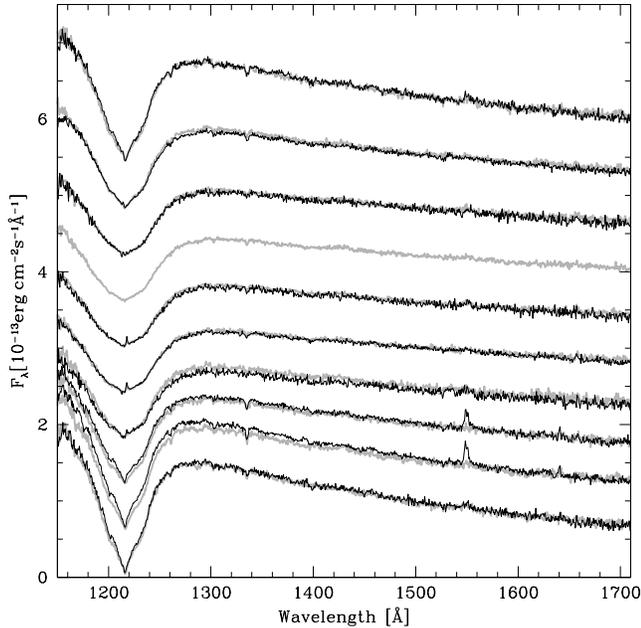}
\caption{Comparison between the \textit{HST}/STIS low state spectra of
  AM\,Her obtained in July 2002 STIS (gray lines) and the November
  2003 data (black lines). Phases and flux offsets of both sets of
  STIS spectra are identical, and are detailed in
  Fig.\,\ref{f-phase_spec}. The November 2003 observations did not
  have as complete a phase coverage as the July 2002 observations,
  which explains the lack of a spectrum for phase bin centered on
  phase 0.65.}
\end{figure}

\section{Comparison of the different epochs of the FUSE and STIS
  observations}
\label{s-comparison}
Figure\,\ref{f-epoch_comp} compares the orbital minimum and maximum
\textit{FUSE} and \textit{HST}/STIS observations of AM\,Her obtained
at different epochs (Fig.\,\ref{f-aavso}, Table\,\ref{t-obslog}).

The two \textit{FUSE} low state data are virtually identical in flux
level and in spectral shape.  The principal differences are that the
September 2002 spectrum was less contaminated by airglow emission and
had --~unfortunately~-- a lower signal to noise as a result of a
shorter exposure time than the May 2002 spectrum. Plotted along with
our own low state data are the June 2002 high state observations
obtained by \citet{hutchingsetal02-1} just after AM\,Her returned from
a short-lived low state. The 1100--1150\,\AA\ continuum fluxes during
the faint phase and the bright phase differ by factors $\sim2.0$ and
$\sim1.4$, respectively. Obviously, the ongoing accretion during the
high state results in emission lines, primarily
\Lines{C}{III}{977,1175}, \Line{N}{III}{992}, \Lines{O}{VI}{1032,1038}
(see \citealt{hutchingsetal02-1} for a study of the emission
lines). It is interesting to note that the spectral shape of the high
state faint phase spectrum does not differ strongly from the low state
faint phase spectrum, subtracting a constant flux of
$\sim0.4\times10^{-13}$\,\ecsa\ brings the high state and low state
spectra to close agreement, suggesting that the unheated white dwarf
was still the dominant FUV flux contribution at that time, and that
accretion adds a featureless nearly flat continuum component. During
the bright phase, a broad \Lb\ absorption from the white dwarf
photosphere is still clearly visible in the high state spectrum, but
the spectral shape is markedly different from the low state. In
particular, substantial flux is observed at $\lambda\la970$\,\AA\
during the high state, rising towards the Lyman edge, where
practically no continuum flux was observed during the low state. This
additional flux component is a mixture of emission from the heated
polar cap (now hotter than during the low state) and of emission from
the accretion stream/funnel.  

As for the two sets of \textit{FUSE} low state observations, also the
July 2002 and the November 2003 \textit{HST}/STIS low state data are
nearly identical at all orbital phases, except for the appearance of
some weak \Line{C}{IV}{1550} emission near $\porb\simeq0.15-0.25$,
which is most likely caused by an intermittent accretion event. On
close inspection, it appears that the continuum flux during that
episode is slightly elevated over that of the July 2002 data. As
mentioned in Sect.\,\ref{s-obshst}, the November 2003 STIS data were
obtained just during the rise to an intermediate optical brightness,
and it is possible that this observation captured the onset of
accretion.

\section{Discussion}
\label{s-discussion} During the low state, the FUV continuum flux of
AM\,Her is entirely made of emission from the white dwarf and a
moderately hot polar cap covering $\sim10$\% of the white dwarf
surface. With our \textit{FUSE} and \textit{HST}/STIS observations we
have probed the low state FUV emission of AM\,Her at four different
epochs, and detected practically no variation of the continuum flux at
any orbital phase. This finding implies that neither the flux
contribution of the white dwarf, nor that of the hot spot vary as a
function of the time spent in a low state. In other words, the
temperature of the white dwarf as well as the temperature, size, and
location of the hot pole cap remain constant over periods of several
months. This confirms the results of \citet{gaensickeetal95-1}, who
did not find any significant change in the white dwarf and hot spot
parameters when analyzing \textit{IUE} low state spectroscopy obtained
at five different epochs, however, our \textit{FUSE} and
\textit{HST}/STIS results provide a much tighter constraint on the
absence of changes in the white dwarf and hot spot parameters. 

This finding is somewhat counter-intuitive, as FUV observations of
dwarf novae clearly showed a short-term response of the white dwarf to
changes in the accretion rate. In those systems, the white dwarf is
heated during dwarf nova outbursts, and subsequently cools
exponentially to its quiescence temperature. The post-outburst cooling
time scales that have been found range from a few days to a few weeks
(comparable to the duration of the outburst itself, e.g. VW\,Hyi:
\citealt{gaensicke+beuermann96-1, sionetal96-1}) to many years (much
longer than the duration of the outburst, e.g.  e.g. WZ\,Sge:
\citealt{slevinskyetal99-1, godonetal04-1, longetal04-1}; or AL\,Com,
\citealt{szkodyetal03-1}). From the observations presented here and by
\citet{gaensickeetal95-1}, we conclude that the cooling of the white
dwarf upon the transition from a high state to a low state proceeds
either on time scales of a few weeks or less (so that it has not been
caught by any of the past FUV observations) or on time scales much
longer than the duration of a low state (so that no noticeable cooling
is observed throughout the low state), or that the bulk of the white
dwarf is not heated at all during the high state.

Assuming that the white dwarf temperature is governed by accretion, we
use Figure\,1 from \citet{townsley+bildsten03-1} to estimate the
secular mean accretion rate of AM\,Her, and find $\dot
M=3\times10^{-10}$\,\msy. This value can be compared with $\dot
M=1.2\times10^{-10}$\,\msy, estimated by \citet{hessmanetal00-1} from
21 years of the long-term optical light curve of AM\,Her together with
a magnitude-dependent bolometric correction. The agreement within a
factor 2.5 is fairly satisfying, considering the uncertainties in the
bolometric correction used by \citet{hessmanetal00-1}, and the fact
that the temperatures predicted by \citet{townsley+bildsten03-1}
reflect the mean accretion rate averaged over much longer time
scales. 

A rather puzzling result is that the temperature and the size of the
pole cap apparently do not vary much, if at all, during the low
state. Two plausible options explaining this finding is either that
the pole cap is kept hot by ongoing accretion at a low level, as
suggested by \citet{gaensickeetal95-1} on the basis that both hard
X-ray emission and cyclotron radiation are detected occasionally
during low states, or that the pole cap is sufficiently deep heated by
accretion to remain at constant temperatures for the duration of a low
state.

The low state luminosity of the hot pole cap is computed as the sum
over all surface elements
\begin{equation}
 L_\mathrm{cap}=\sum \sigma A_\mathrm{SE} (T_\mathrm{SE}^4-\Twd^4)=1.34\times10^{31}\es
\end{equation}
where $\sigma$ is the Stefan-Boltzmann constant, $A_\mathrm{SE}$ the
area of each individual surface element, and $T_\mathrm{SE}$ its
temperature. From $L_\mathrm{acc}=G \Mwd\dot M/\Rwd$, and
assuming $\Mwd=0.78$\,\Msun and $\Rwd=7.61\times10^8$\,cm, we obtain
$1.23\times10^{14}\mathrm{gs^{-1}}=1.54\times10^{-12}\msy$. Following
\citet{gaensickeetal95-1}, roughly equal parts of accretion luminosity
are expected in hard X-ray emission, cyclotron radiation, and FUV
emission from the pole cap, adding up to a low state accretion rate of
$\simeq6\times10^{-12}$\,\msy. Such a low level of mass transfer is
not implausible, as it just implies that the secondary star atmosphere
did not fully withdraw from the $L_1$ point. From the observations,
there is evidence that X-ray, optical, and infrared activity during
the low state is variable \citep[e.g.][]{demartinoetal98-3,
kafkaetal05-1, baileyetal91-1}, but it is not clear if non-stationary
accretion could maintain the pole cap at the observed constant
temperature.

The alternative explanation is deep heating during the high state.
\citet{gaensickeetal99-1} showed that the white dwarf in the VY\,Scl
star TT\,Ari remained at $\simeq40\,000$\,K for an entire year during
a prolonged low state, in that case deep heating is clearly the only
explanation as the white dwarf temperature is much too hot to be
explained by ongoing accretion during the low state. Our analysis of
the high state \citep{gaensickeetal98-2} and the low state (this
paper) suggest, however, that the spot moves somewhat in co-latitude
as a funtion of accretion rate, which contrasts with the idea of a deeply
heated hot spot at a fixed location. Moreoever, it is a challenge to
stellar structure theory to test whether or not such a substantial
temperature inhomogeneity can remain in place for the observed periods
of several months. 

Additional observational input into testing both hypotheses for the
origin of the low state hot spot would be an intense FUV monitoring of
AM\,Her during both high states and low states, and most importantly,
during the transitions from between the two states, to accurately
track the evolution of the pole cap in temperature, size, and
extension. While AM\,Her is by far the best studied case of a heated
pole cap on a white dwarf, large heated pole caps have been identified
in more than half a dozen other polars, suggesting that they are a
fundamental feature of these systems (e.g. \citealt{schwope90-1,
stockmanetal94-1, ferrarioetal96-1, gaensicke99-1, gaensickeetal00-1,
rosenetal01-1, araujo-betancoretal05-2}).

\section{Summary}
FUV \textit{FUSE} and \textit{HST}/STIS spectroscopy of the
prototypical polar AM\,Her obtained during the low state shows that
the white dwarf is the dominant source of emission. The data reveal a
strong orbital modulation of the FUV flux as well as a strong
variation of the Lyman absorption lines from the white dwarf
photosphere. A white dwarf of 19\,400\,K with a hot pole cap of
$\simeq34\,000-40\,000$\,K covering $\sim10$\% of the white dwarf
qualitatively fits the orbital phase-resolved \textit{FUSE} and STIS
spectra. Using \citeauthor{thorstensen03-1}'s astrometric distance of
AM\,Her and a \citet{wood95-1} mass-radius relation, we determine the
white dwarf mass to be $\Mwd=0.78\pm{0.12\atop0.17}$\,\Msun. The
absence of any noticeable absorption intrinsic to AM\,Her, other than
the Lyman lines, suggests very low average metal abundances in the
white dwarf atmosphere.  Based on four FUV low-state
observations of AM\,Her obtained at different epochs it appears that
the location, temperature, and size of the hot pole cap do not vary as
a function of the time spent in a low state. Comparing our low state
results with those obtained from an \textit{HST}/GHRS observation
carried out during a high state, we find some evidence for a small
change in co-latitude of the spot, with the spot being located closer
to the magnetic pole during the low state than during the high state.
Detailed monitoring observations covering the transition between high and low
state would be important to determine the time scale on which the pole
cap adjusts from the high state to the low state parameter. In
addition, it would be interesting to compare the structure of the
cyclotron emitting region via a a Stokes imaging analysis
\citep{potteretal98-1}.

\acknowledgements{We thank Susanne Friedrich for supplying tables with
  the wavelengths of the Lyman Zeeman components. BTG was supported by
  a PPARC Advanced Fellowship.  Additional support was provided
  through NASA grant GO-9635 from the Space Telescope Science
  Institute, which is operated by the Association of Universities for
  Research in Astronomy, Inc., under NASA contract NAS 5-26555.  We
  greatly acknowledge the variable star observations from the AAVSO
  International Database contributed by observers worldwide and used
  in this research. We thank the referee, John Thorstensen, for his
  useful comments.}

\bibliographystyle{apj}
\bibliography{aamnem99,aabib}

\end{document}